\def\papertitle{Enhancing Automatic Chord Recognition via Pseudo-Labeling and Knowledge Distillation}
\def\paperauthorA{Nghia Phan, Rong Jin}
\def\paperauthorB{Gang Liu}
\def\paperauthorC{Xiao Dong}

\documentclass[twoside,a4paper]{article}
\usepackage{textcomp}

\usepackage[print]{dafx26v3}
\usepackage{amsmath,amssymb,amsfonts,amsthm}
\usepackage{siunitx}
\usepackage{euscript}
\usepackage[T1]{fontenc}
\usepackage[utf8]{inputenc}
\usepackage{ifpdf}
\usepackage[english]{babel}
\usepackage{caption}
\usepackage{subfig} % or can use subcaption package
\usepackage{color}
\usepackage{xcolor}
\usepackage{booktabs}
\usepackage{colortbl}
\usepackage{multirow}
\usepackage{algorithm}
\usepackage{algorithmic}
\usepackage{url}
\usepackage{microtype} % Improves line breaking and reduces overfull hboxes

\definecolor{btcblue}{RGB}{0,120,200}
\definecolor{e1dgreen}{RGB}{0,150,80}
\definecolor{btcbluehighlight}{RGB}{223,239,250}
\definecolor{e1dgreenhighlight}{RGB}{221,243,225}

\def\KL{{\mathrm{KL}}}
\def\CE{{\mathrm{CE}}}

\def\WCSR{{\mathrm{WCSR}}}
\def\ACQA{{\mathrm{ACQA}}}
\def\CSR{{\mathrm{CSR}}}

\input glyphtounicode
\ninept

\newcounter{numauth}
\newcounter{listcnt}
\newcommand\authcnt[1]{\ifdefined#1 \stepcounter{numauth} \fi}

\newcommand\addauth[1]{
\ifdefined#1 
\stepcounter{listcnt}
\ifnum \value{listcnt}<\value{numauth}
\appto\authorslist{, #1}
\else
\appto\authorslist{~and~#1}
\fi
\fi}
\authcnt{\paperauthorB}
\authcnt{\paperauthorC}
\authcnt{\paperauthorD}
\authcnt{\paperauthorE}
\authcnt{\paperauthorF}
\authcnt{\paperauthorG}
\authcnt{\paperauthorH}
\authcnt{\paperauthorI}
\authcnt{\paperauthorJ}
\def\authorslist{\paperauthorA}
\addauth{\paperauthorB}
\addauth{\paperauthorC}
\addauth{\paperauthorD}
\addauth{\paperauthorE}
\addauth{\paperauthorF}
\addauth{\paperauthorG}
\addauth{\paperauthorH}
\addauth{\paperauthorI}
\addauth{\paperauthorJ}
\usepackage{times}
\newif\ifpdf
\ifx\pdfoutput\relax
\else
   \ifcase\pdfoutput
      \pdffalse
   \else
      \pdftrue
   \fi
\fi

\ifpdf % compiling with pdflatex
  \usepackage[pdftex,
    pdftitle={\papertitle},
    pdfauthor={\authorslist},
    pdfsubject={Proceedings of the 29th International Conference on Digital Audio Effects (DAFx26)},
    colorlinks=false, % links are activated as color boxes instead of color text
    bookmarksnumbered, % use section numbers with bookmarks
    pdfstartview=XYZ % start with zoom=100% instead of full screen; especially useful if working with a big screen :-)
  ]{hyperref}
  \usepackage[pdftex]{graphicx}
\else % compiling with latex
  \usepackage[dvips]{epsfig,graphicx}
  \usepackage[dvips,
    pdftitle={\papertitle},
    pdfauthor={\authorslist},
    pdfsubject={Proceedings of the 29th International Conference on Digital Audio Effects (DAFx26)},
    colorlinks=false, % no color links
    bookmarksnumbered, % use section numbers with bookmarks
    pdfstartview=XYZ % start with zoom=100% instead of full screen
  ]{hyperref}
\fi
\usepackage[hypcap=true]{caption}
\title{\papertitle}

\affiliation
  {{\it Nghia Phan}$^{1}$, {\it Rong Jin}$^{1}$, {\it Gang Liu}$^{2}$, and {\it Xiao Dong}$^{3}$}
  {$^{1}$California State University, Fullerton, CA, USA \ \ 
   $^{2}$Microsoft, Redmond, WA, USA \ \
   $^{3}$Beijing Normal-Hong Kong Baptist University, China \\
   {\footnotesize \tt ptnghia@csu.fullerton.edu, rong.jin@fullerton.edu, liu.gang@microsoft.com, xiaodong@bnbu.edu.cn}
  }
\begin{document}
% more pdf-tex settings:
\ifpdf % used graphic file format for pdflatex
  \DeclareGraphicsExtensions{.png,.jpg,.pdf}
\else  % used graphic file format for latex
  \DeclareGraphicsExtensions{.eps}
\fi

%\makeatletter
%\pdfbookmark[0]{\@pdftitle}{title}
%\makeatother

\maketitle

\begin{abstract}
Automatic Chord Recognition (ACR) is constrained by the scarcity of aligned chord annotations, which are costly to acquire. At the same time, open-weight pre-trained models are more accessible than their proprietary training data. In this work, we present a two-stage training pipeline that leverages pre-trained models together with unlabeled audio. The proposed method decouples training into two stages. In the first stage, we use the pre-trained BTC model \cite{Park19} as a teacher to generate pseudo-labels for over 1,000 hours of diverse unlabeled audio and train a student model solely on these pseudo-labels. In the second stage, the student is continually trained on ground-truth labels as they become available. To prevent catastrophic forgetting of the representations learned in the first stage, we apply selective knowledge distillation (KD) from the teacher as a regularizer. In our experiments, two models (BTC, 2E1D) were used as students. In Stage 1, using only pseudo-labels, the BTC student achieves about 99\% of the teacher's performance, while the 2E1D model achieves about 97\% of the teacher's performance across seven standard \texttt{mir\_eval} metrics. After continual training with labeled data in Stage 2, the resulting BTC student model consistently surpasses both the traditional supervised learning baseline and the original pre-trained teacher model across all metrics. The resulting 2E1D student model also outperforms the supervised baseline and approaches teacher-level performance, with both models demonstrating substantial gains on rare chord qualities.
\end{abstract}

\section{Introduction}
\label{sec:intro}

% REVISED VERSION
Automatic Chord Recognition (ACR) from audio signals remains a fundamental task in Music Information Retrieval (MIR) that aims to identify the harmonic content of audio recordings by outputting a sequence of chord labels. While large-scale labeled datasets are readily available for many machine learning domains, ACR faces significant data constraints: publicly available labeled chord datasets remain limited in both size and diversity \cite{Humphrey15}. This scarcity stems from the substantial manual effort required for precise audio-label alignment and consistent harmonic interpretation, as chord boundaries are inherently ambiguous and context-dependent \cite{Harte10,Pauwels19}. Furthermore, chord vocabularies are large and highly imbalanced; models excel on frequent major/minor chords but underperform on rare seventh and extended chords \cite{Humphrey15,Pauwels19,Bortolozzo21}. 

Recent semi-supervised ACR methods \cite{Bortolozzo21,Li23_contrastive} have attempted to integrate unlabeled data into training. They use pseudo-labeling or contrastive learning but require ground-truth labels from the outset to fuse labeled and unlabeled data within a single training run. This tight coupling limits their utility when labeled data is initially unavailable or when training data remains proprietary.

To overcome these limitations, we experiment with two semi-supervised learning paradigms: \textit{pseudo-labeling}~\cite{Lee13} and \textit{knowledge distillation} (KD)~\cite{Hinton14}. Pseudo-labeling is a self-training technique where a pre-trained ``teacher'' model generates hard or soft target predictions on unlabeled audio, which serve as synthetic ground truth to train a ``student'' network~\cite{Lee13}. Knowledge distillation is a model regularization and compression framework wherein a student is optimized to match the full output probability distribution (soft logits) of a teacher model, transferring knowledge about inter-class relationships that hard annotations omit~\cite{Hinton14}. By combining these paradigms in a decoupled, two-stage pipeline (see Figure~\ref{fig:training_pipeline}), we train a student model that can match or exceed the teacher's capabilities. In the first stage, a pre-trained teacher model generates pseudo-labels for over 1,000 hours of diverse unlabeled audio, and a student model is trained solely on these pseudo-labels until convergence, without requiring any ground-truth annotations. In the second stage, when labeled data becomes available, the pseudo-label-trained student is continually trained on ground-truth labels. To retain the representations acquired in the first stage, we apply selective KD from the teacher as a regularizer throughout the second stage.

Our two-stage training approach decouples labeled and unlabeled data, making the training viable even when labeled data is initially unavailable. When data collection is expensive, this offers a practical alternative by training a model entirely without labels at only a minor accuracy cost. Notably, open-weight pre-trained models are often more readily available than their training data, offering a practical way to leverage teacher knowledge without access to proprietary datasets. Additionally, we show that the accuracy gains are disproportionately concentrated on rare chord qualities. 

We also introduce a compact, dual-encoder architecture (2E1D) that is based on the Transformer architecture~\cite{Vaswani17} and is lighter-weight than the teacher model. We demonstrate that our method can generalize across architectures. Our experiments show that the best resulting student surpasses both the traditional supervised learning approach and the pre-trained teacher model across all standard \texttt{mir\_eval} metrics \cite{raffel14}, with particularly large gains on rare chord qualities. We open-source a chord-recognition web application\footnote{\url{https://github.com/ptnghia-j/ChordMiniApp}} that lets practitioners run and validate our models on their own audio locally.

\section{Related Work}
\label{sec:related}

% \subsection{Pseudo-Labeling in Semi-Supervised Learning}

Pseudo-labeling, where a trained model generates labels for unlabeled data, has become a cornerstone of semi-supervised learning~\cite{Lee13}. The Noisy Student framework~\cite{Xie20} demonstrated that iteratively training larger student models on pseudo-labeled data with noise injection can surpass teacher performance, establishing a paradigm for leveraging unlabeled data at scale. FixMatch~\cite{Sohn20} combined consistency regularization with pseudo-labeling using confidence thresholds, while Mean Teacher~\cite{Tarvainen17} maintained an exponential moving average of student weights to produce consistency targets. Meta Pseudo Labels~\cite{Pham21} further improved teacher--student training by jointly optimizing the teacher based on student feedback. Within audio signal processing, pseudo-labeling has been successfully applied to various tasks including speech recognition~\cite{Likhomanenko23}, piano transcription~\cite{Strahl24}, and music tagging~\cite{Hung2023}. For ACR specifically, Bortolozzo et al.~\cite{Bortolozzo21} adapted Noisy Student for rare chord recognition, employing confidence filtering and iterative teacher--student training to address class imbalance. Li et al.~\cite{Li23_contrastive} combined contrastive pre-training with noisy-student semi-supervision for large-vocabulary chord recognition. However, these approaches typically require ground-truth labels from the outset and train on mixed pseudo-label and ground-truth signals within a single pipeline. In contrast, we investigate whether \emph{separate} incremental learning stages can achieve comparable or superior performance by first training on pseudo-labels alone, then adapting to labeled data.

% \subsection{Knowledge Distillation as Regularization}

Transferring knowledge via temperature-scaled soft targets, or knowledge distillation~\cite{Hinton14}, has been shown to provide richer supervision than hard labels. Yuan et al.~\cite{Yuan23} studied conditions under which biased soft labels can still improve student generalization. Chen~\cite{Chen21_loss} showed that KD can be interpreted as a form of output regularization, while Mansourian et al.~\cite{mansourian2025comprehensive} detailed its general regularizing effects. In continual learning, KD has been used to preserve prior knowledge while adapting to new data, mitigating catastrophic forgetting~\cite{LiHoiem18, French99}. Learning without Forgetting (LwF)~\cite{LiHoiem18} pioneered using distillation from the model's own previous state to retain old knowledge when learning new tasks. KD has primarily been used for model compression, training smaller, more efficient students~\cite{Hung2023}. In real-time audio systems, such compression is vital for deploying computationally demanding deep learning models onto low-latency digital audio workstations (DAWs) or embedded hardware targets, as explored in recent studies on real-time inference engines~\cite{Stefani22} and structured network pruning~\cite{Clarke25}. While we similarly optimize a compact student model (2E1D), we apply KD differently: as a regularization mechanism during continual learning that anchors the student to the teacher's generalized representations. Consequently, KD enables the student to adapt to new labels while retaining prior pseudo-label knowledge, reducing catastrophic forgetting.

% \subsection{Continual Learning in Music Information Retrieval}

Continual learning addresses the challenge of learning from non-sta\-tion\-ary data streams without forgetting previously acquired knowl\-edge~\cite{vandeVen19}. In the MIR domain, this challenge is particularly relevant given the ongoing creation of new music and evolving annotation standards. While task-incremental and class-incremental scenarios have received significant attention, data-incremental continual learning remains underexplored for ACR. In this setting, the task and label space remain unchanged but new data arrive.

% Our work bridges these three areas by proposing a two-stage pipeline: (1) initial training on large-scale pseudo-labeled data without any ground-truth supervision, and (2) data-incremental continual learning when labeled data becomes available, using KD as a regularizer to preserve the teacher's generalized representations.

\section{Methodology}
\label{sec:methodology}

\subsection{Problem Formulation}
Let $\mathcal{D}_l = \{(x_i, y_i)\}_{i=1}^{N_l}$ denote a small labeled dataset where $x_i \in \mathbb{R}^{T \times F}$ represents time–frequency features (e.g., Constant-Q Transform) with $T$ frames and $F$ frequency bins, and $y_i \in \{1, 2, \ldots, V\}^T$ are frame-wise chord labels over a vocabulary of size $V$. Let $\mathcal{D}_u = \{x_j\}_{j=1}^{N_u}$ represent large-scale unlabeled datasets where $N_u \gg N_l$. Our objective is to train an effective student model $f_s$ by leveraging pseudo-labels generated from $\mathcal{D}_u$. This mitigates reliance on expensive manual annotations while maintaining competitive performance.

\subsection{Training Method}
We employ a pre-trained teacher model $f_t: \mathbb{R}^{T \times F} \rightarrow \mathbb{R}^{T \times V}$ to generate pseudo-labels for unlabeled data. Any open-weight ACR model can serve as the teacher. In our experiments, we use the BTC model from Park et al.~\cite{Park19} as the teacher model and adopt the standard vocabulary size $V=170$ for all training settings. For each unlabeled sequence $x_j \in \mathcal{D}_u$, pseudo-labels are generated via frame-wise argmax over teacher outputs:
\begin{equation}
\hat{y}_j^{(n)} = \arg\max_{c \in \{1,\ldots,V\}} \bigl[f_t(x_j)\bigr]_{n,c}, 
\quad n=1,\ldots,T.
\end{equation}
To preserve temporal coherence, we keep sequence boundaries intact during teacher inference (e.g., padding only at segment ends) and convert 100\% of frames to pseudo-labels without confidence filtering, yielding a pseudo-labeled dataset:
\begin{equation}
\mathcal{D}_u^{(p)} = \{(x_j, \hat{y}_j)\}_{j=1}^{N_u}.
\end{equation}

Our method uses $M$ complementary unlabeled datasets where $\mathcal{D}_u = \mathcal{D}_1 \cup \mathcal{D}_2 \cup \cdots \cup \mathcal{D}_M$. The specific datasets used in our experiments are described in Section~\ref{sec:data}.

\begin{figure}[htb]
    \centering
    \includegraphics[width=0.45\textwidth]{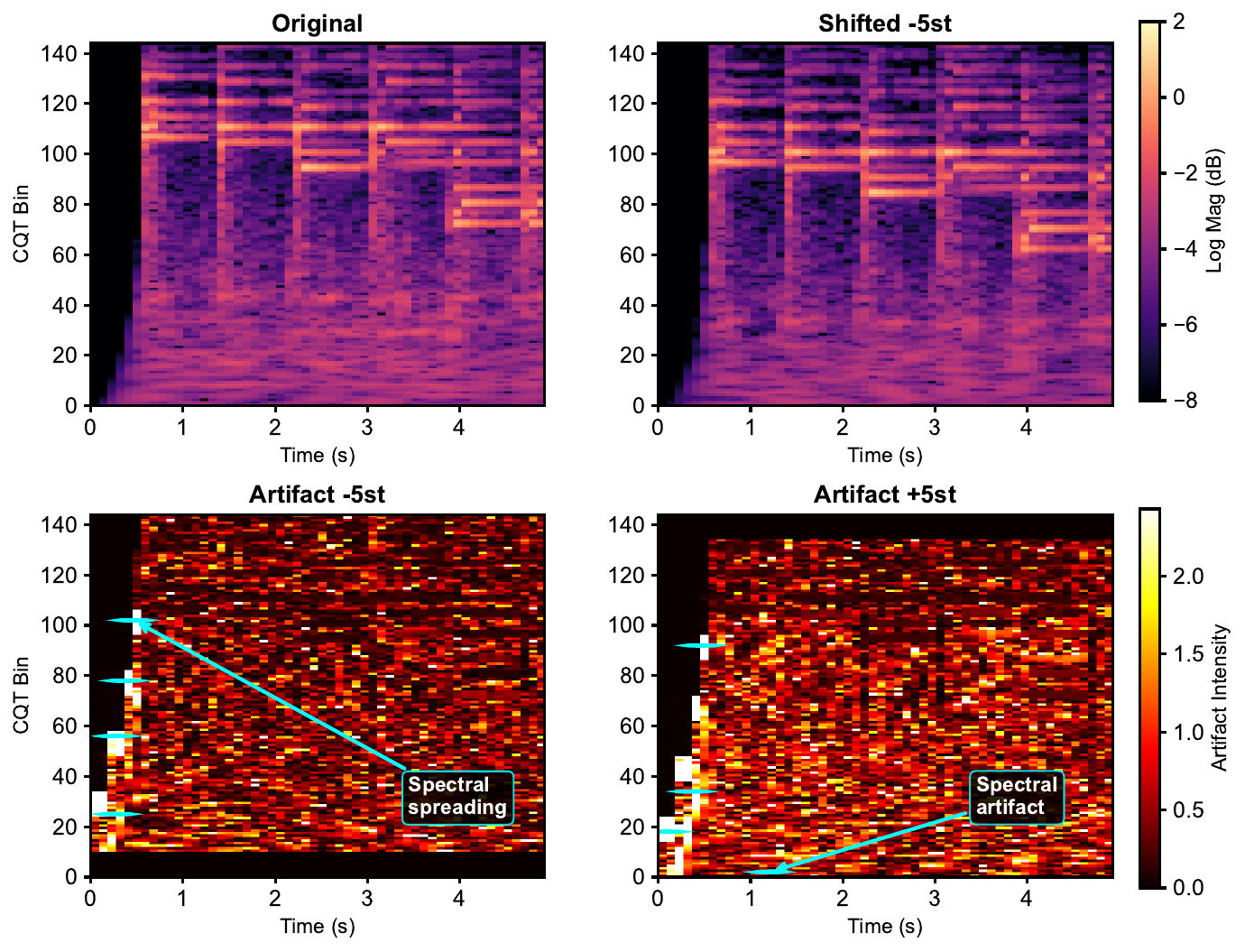}
    \caption{Constant-Q Transform (CQT) comparison revealing pitch-shifting artifacts. Annotations indicate spectral spreading (low-frequency energy diffusion) and spectral artifacts (high-frequency noise) introduced by the phase-vocoder shift.}
    \label{fig:pitch_artifacts}
\end{figure}

\subsubsection{Data Augmentation vs.\ Natural Chord Root Coverage}
\label{sec:augmentation}

The supervised chord recognition task typically requires pitch-shifting data augmentation to address severe chord root imbalance in labeled datasets. Studies show that widely used datasets exhibit strong biases toward C and G major keys, comprising a substantial portion of the total duration, while keys typically notated with many accidentals (e.g., F$\sharp$, D$\flat$) are underrepresented~\cite{Humphrey15}. Without augmentation, models overfit to key-specific spectral patterns and fail to generalize chord templates across all 12 pitch classes. Recent semi-supervised methods similarly rely on pitch-shifting when training on the labeled subset. However, pitch-shifting introduces audible artifacts. Phase-vocoder-based methods, including the Rubber Band library commonly used for music augmentation, produce transient smearing, phasiness, and spectral discontinuities~\cite{Driedger16}. These artifacts manifest as temporal blurring in Constant-Q Transform (CQT) features and artificial energy spreading across frequency bins (Figure~\ref{fig:pitch_artifacts}).

% We observe that large-scale pseudo-labeling provides an alternative that naturally eliminates the need for pitch-shifting. We analyzed chord root distributions across 101,575 pseudo-labeled tracks ($\sim$1,300 hours) from FMA, DALI, and MAESTRO. As shown in Figure~\ref{fig:key_distribution}, all 12 pitch classes are well-represented in accumulated root note distribution. The Shannon entropy reaches 3.53 bits (the maximum being 3.58 bits), corresponding to 98.4\% uniformity. This natural coverage arises because diverse unlabeled corpora span multiple genres, artists, and production contexts, each contributing different key preferences that aggregate to near-uniform distribution. Consequently, pseudo-label pretraining exposes the model to patterns in all keys without requiring pitch-shifting, establishing key-invariant representations. In our experiment, applying pitch-shifting during Stage 2 ground-truth fine-tuning shows no improvement for our pre-trained models (Section~\ref{sec:results}), confirming that Stage 1 pre-training already builds key-invariant representations. This allows us to completely omit signal manipulation and its associated artifacts throughout our training pipeline.

\begin{figure}[htb]
    \centering
    \includegraphics[width=0.43\textwidth]{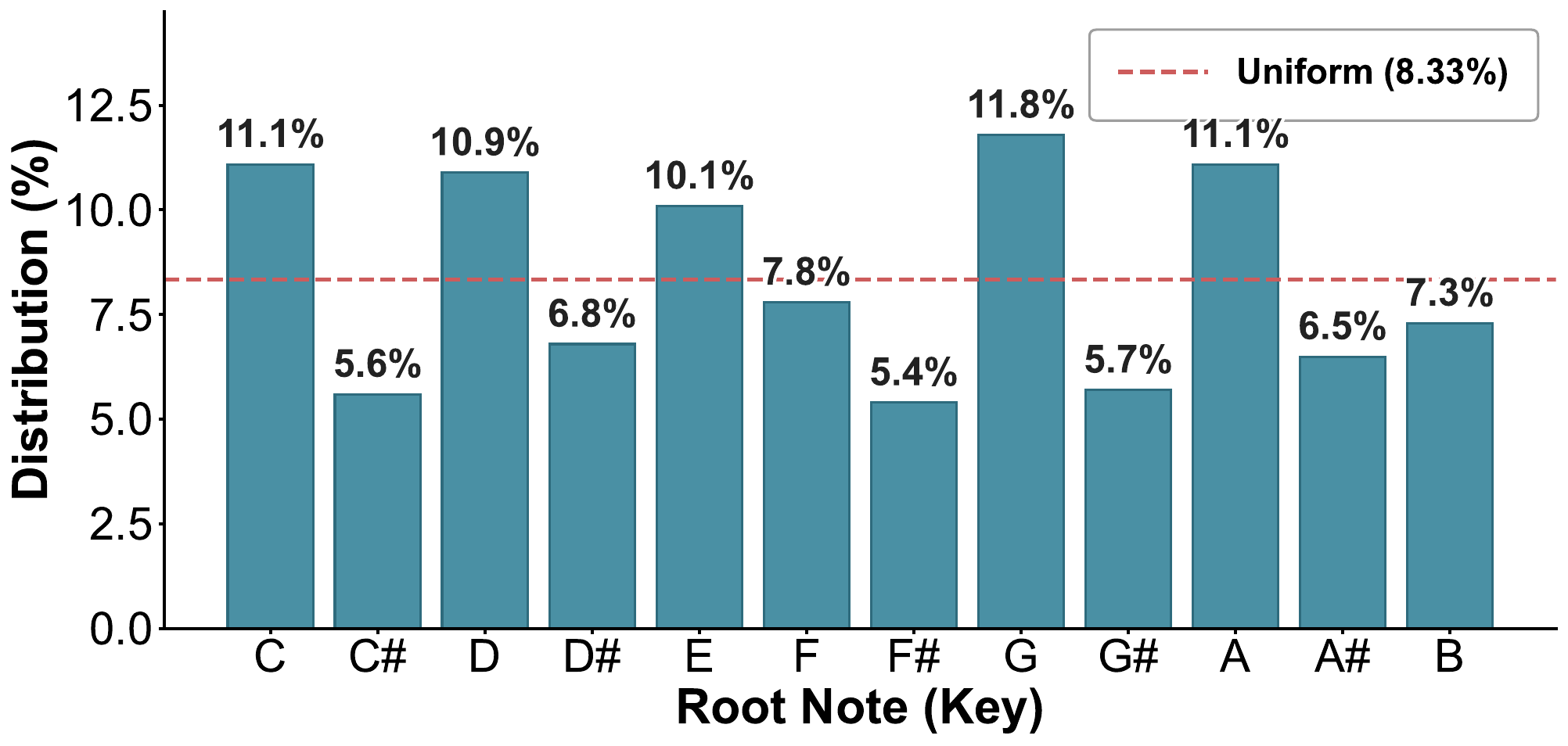}
    \caption{Duration-weighted chord root distribution across pseudo-labeled datasets (Section~\ref{sec:data}). The dashed line indicates uniform distribution (8.33\%). Pitch classes are well-represented with 98.4\% uniformity.}
    \label{fig:key_distribution}
\end{figure}

We observe that large-scale pseudo-labeling provides an alternative that naturally eliminates the need for pitch-shifting. We analyzed chord root distributions across 101,575 pseudo-labeled tracks ($\sim$1,300 hours) from FMA, DALI, and MAESTRO. As shown in Figure~\ref{fig:key_distribution}, all 12 pitch classes are well-represented in the accumulated root note distribution. The Shannon entropy ($H = -\sum_{i=1}^{12} p_i \log_2 p_i$) reaches 3.53 bits (the maximum being $\log_2(12)$ $\approx 3.58$ bits), corresponding to 98.4\% uniformity ($3.53 / \log_2(12) \approx 98.4\%$). This natural coverage arises because diverse unlabeled corpora span multiple genres, artists, and production contexts, each contributing different key preferences that aggregate to a near-uniform distribution. Consequently, pseudo-label pre-training exposes the model to patterns in all keys without requiring pitch-shifting, establishing key-invariant representations. In our experiments, applying pitch-shifting during Stage 2 ground-truth fine-tuning shows no improvement for our pre-trained models (Section~\ref{sec:results}), confirming that Stage 1 pre-training already builds key-invariant representations. This allows us to completely omit signal manipulation and its associated artifacts throughout our training pipeline.

\begin{figure}[htb]
\centering
\includegraphics[width=0.48\textwidth]{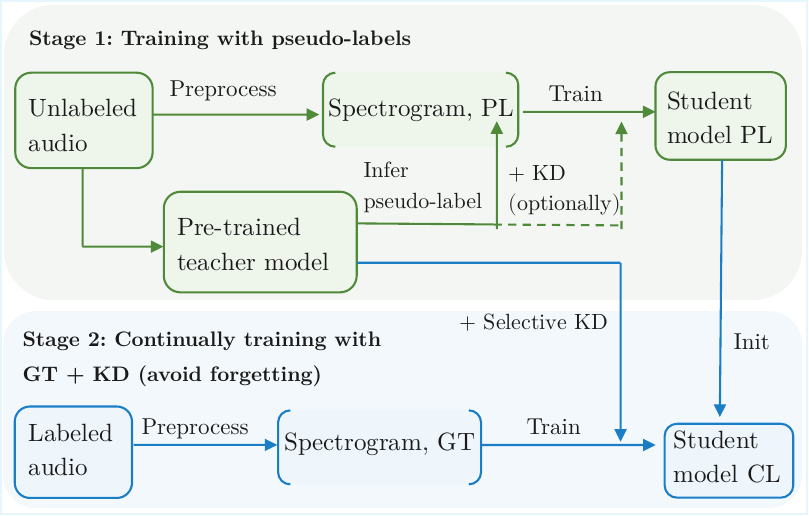}
\caption{Illustration of the proposed two-stage training pipeline. Details of audio data are described in Section~\ref{sec:data}. Note: The resulting Student model CL from Stage 2 can be continually trained when additional labeled data is available.}
\label{fig:training_pipeline}
\end{figure}

\subsubsection{The Training Pipeline}

Figure~\ref{fig:training_pipeline} illustrates the complete two-stage training pipeline. In Stage 1, unlabeled data is first preprocessed into Constant-Q Transform spectrograms following the procedure described in Section~\ref{sec:data}. A pre-trained model is used as a teacher to infer frame-wise pseudo-labels (PL) on the unlabeled audio for each track, producing paired data $(\text{Spectrogram}, \text{PL})$. A student model is then trained on these pseudo-labeled pairs until convergence. Optionally, knowledge distillation (KD) from the teacher's soft targets can be applied during this stage to accelerate convergence; this optional path is depicted as a dashed line in the figure. We denote the resulting model after the first stage of training as the \textit{Student model PL}. Stage 2 begins when the labeled data becomes available. Labeled audio undergoes the same preprocessing, yielding paired data $(\text{spectrogram}, \text{GT})$, where GT denotes ground-truth chord annotations. In this stage, \textit{Student model PL} serves as the weight initialization for a new model called \textit{Student model CL}. Then the \textit{Student model CL} serves as the initialization for subsequent rounds of continual learning as additional labeled datasets are acquired. The selective KD signal (Section~\ref{sec:selectiveKD}) from the original pre-trained teacher is applied throughout the training of \textit{Student model CL}. KD acts as regularization, anchoring the student to the teacher’s distributional knowledge when ground-truth labels conflict with teacher predictions, while still allowing adaptation when they agree. When KD is applied, the training loss is a weighted sum of a classification term and a KD term as follows:

\begin{equation}
\mathcal{L}_{\text{total}} = \alpha \mathcal{L}_{\text{KD}} + (1-\alpha) \mathcal{L}_{C},
\label{eq:total_loss}
\end{equation}
where $\alpha \in [0,1]$ controls the balance between teacher regularization and label supervision, and $\mathcal{L}_{C}$ denotes the classification loss. Specifically:

\begin{equation}
    \mathcal{L}_{C} = 
    \begin{cases} 
    \mathcal{L}_{\mathrm{PL}} = \frac{1}{|\mathcal{D}_u^{(p)}|} \sum_{(x,\hat{y}) \in \mathcal{D}_u^{(p)}} \ell_{\CE}(f_s(x), \hat{y}) & \text{if Stage 1} \\[2ex]
    \mathcal{L}_{\CE} = \frac{1}{|\mathcal{D}_l|} \sum_{(x,y) \in \mathcal{D}_l} \ell_{\CE}(f_s(x), y) & \text{if Stage 2}
    \end{cases},
\end{equation}
where $\mathcal{D}_u^{(p)}$ denotes the pseudo-labeled unlabeled data and $\mathcal{D}_l$ denotes the ground-truth labeled data. Setting $\alpha{=}0$ reduces Eq.~\eqref{eq:total_loss} to pure classification training without KD. The per-frame KD loss is:
\begin{equation}
\mathcal{L}_{\text{KD}} = \tau^2 \cdot D_{\KL}\!\left(\sigma\!\left(\frac{\mathbf{z}_t}{\tau}\right) \Big\| \sigma\!\left(\frac{\mathbf{z}_s}{\tau}\right)\right)
\label{eq:kd_loss},
\end{equation}
where $\sigma(\cdot)$ denotes the softmax function, $\tau > 0$ is the temperature parameter that controls the smoothness of probability distributions, $D_{\KL}$ is the Kullback-Leibler divergence, and $\mathbf{z}_s, \mathbf{z}_t \in \mathbb{R}^V$ are the student and teacher logits, respectively.

\subsubsection{KD as Regularization}
We denote the temperature-softened probability distributions as $\mathbf{p}^{(s)} = \sigma(\mathbf{z}_s/\tau)$ and $\mathbf{p}^{(t)} = \sigma(\mathbf{z}_t/\tau)$, with $p^{(s)}_k$ and $p^{(t)}_k$ denoting the $k$-th class probability. The $\tau^2$ scaling ensures gradient magnitudes remain comparable to cross-entropy~\cite{Hinton14}. The KD gradient with respect to student logits is:
\begin{equation}
\nabla_{\mathbf{z}_s} \mathcal{L}_{\text{KD}} = \tau\, (\mathbf{p}^{(s)} - \mathbf{p}^{(t)}).
\label{eq:kd_gradient}
\end{equation}
This gradient ``pulls'' student predictions toward the teacher's distribution. Combining with the classification term from Eq.~\eqref{eq:total_loss}:
\begin{equation}
\frac{\partial \mathcal{L}_{\text{total}}}{\partial z_{s,k}} = (1-\alpha) \frac{\partial \mathcal{L}_{C}}{\partial z_{s,k}} + \alpha \tau \big( p^{(s)}_k - p^{(t)}_k \big).
\label{eq:combined_gradient}
\end{equation}
The second term acts as a regularizer that anchors student predictions to the teacher's distribution. When ground-truth labels conflict with teacher predictions (e.g., due to annotation noise or misalignment), this regularization prevents overfitting to erroneous labels. Conversely, when labels align with teacher predictions, the regularization does not impede adaptation. This property is particularly beneficial for ACR, where annotation inconsistencies are common due to subjective harmonic interpretation~\cite{Harte10,Pauwels19}.

\subsubsection{Selective KD}
\label{sec:selectiveKD}
To further improve robustness, we introduce \textit{selective KD} that filters the teacher signal based on prediction confidence. Let $\gamma$ denote the teacher's maximum temperature-softened probability for a given frame. We define an asymmetric weighting function $w(\gamma) \in [0,1]$:
\begin{equation}
w(\gamma) = \begin{cases}
0 & \text{if } \gamma < \theta_{\min} \\
1 & \text{if } \theta_{\min} \le \gamma \le \theta_{\max} \\
1 - \lambda \cdot \frac{\gamma - \theta_{\max}}{1 - \theta_{\max}} & \text{if } \gamma > \theta_{\max}
\end{cases},
\end{equation}
where $\theta_{\min}$ filters out uninformative low-confidence samples, $\theta_{\max}$ down-weights overconfident predictions that may bias toward majority classes, and the factor $\lambda \in [0,1]$ controls how strongly overconfident predictions are down-weighted. The weighted KD loss becomes $\mathcal{L}_{\text{KD}}^{\text{sel}} = \frac{1}{T}\sum_{n=1}^T w(\gamma_n) \cdot \mathcal{L}_{\text{KD}}^{(n)}$ over all $T$ frames, where $\mathcal{L}_{\text{KD}}^{(n)}$ denotes the per-frame KD loss of Eq.~\eqref{eq:kd_loss} evaluated at frame $n$ and $\gamma_n$ the corresponding teacher confidence. Samples near decision boundaries (moderate confidence) contain valuable uncertainty information, so we preserve full weight in the informative range $[\theta_{\min}, \theta_{\max}]$. We apply selective KD throughout all Stage~2 continual learning experiments to stabilize training by reducing gradient variance from extreme-confidence samples. We set the hyperparameters to $\theta_{\min}=0.1, \theta_{\max}=0.9, \lambda=0.8$ as robust defaults based on the observed teacher-confidence distribution.

\begin{figure}[t!]
    \centering
    \includegraphics[width=0.45\textwidth]{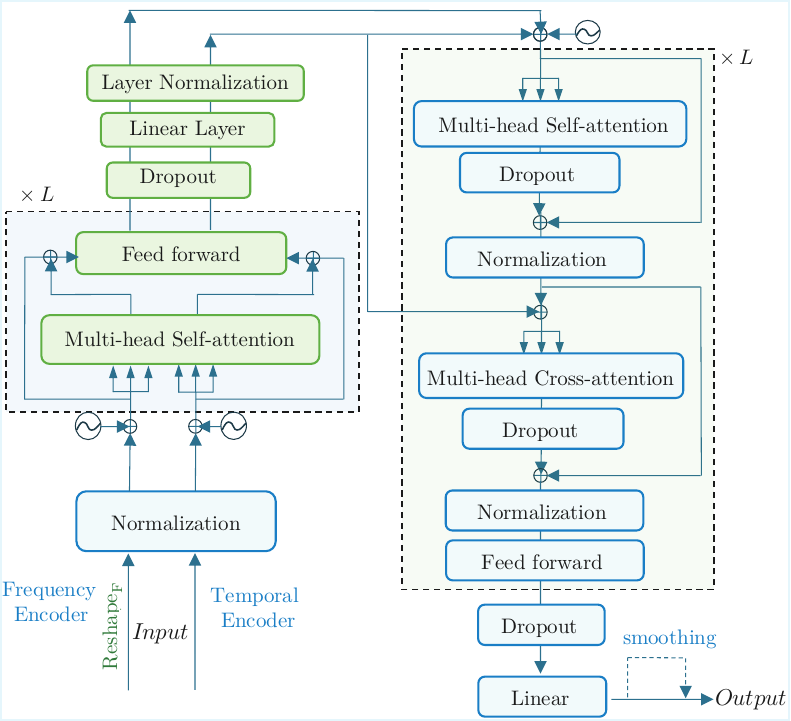}
    \caption{Experimental Dual Encoder Architecture (2E1D): The model consists of separate temporal and frequency encoders that process CQT features independently before fusion for classification.}
    \label{fig:dual_encoder_appendix}
\end{figure}

\subsubsection{Experimental Models}
BTC \cite{Park19} serves as both the pseudo-labeling teacher and the baseline for self-distillation experiments. The model uses a direct frame projection followed by a stack of bi-directional transformer layers with position-wise feed-forward blocks, representing a \emph{deeper} architecture with sequentially stacked attention layers.

For cross-architecture validation, we introduce a new compact dual-encoder architecture (2E1D) \footnote{Model: \url{https://github.com/ptnghia-j/ChordMini}} for chord recognition that is purely transformer-based~\cite{Vaswani17} (without any CNN layer). As shown in Figure~\ref{fig:dual_encoder_appendix}, 2E1D adopts a \emph{wider} design: (1) a frequency encoder that groups CQT bins into spectral clusters and applies self-attention to learn harmonic relationships across frequency bands; (2) a temporal encoder that processes full-band features to model chord progression patterns over time; and (3) a cross-attention fusion block that combines the two streams for final chord classification. At inference time, we apply a temporal smoothing pipeline over output logits to reduce frame-level prediction jitter. Each class logit channel is convolved with a normalized 1D Gaussian kernel $g[n] = \exp(-n^2 / (2\sigma_g^2))/Z$, where $\sigma_g = K/6$ for a kernel of width $K$, normalization factor $Z$, and with replicate padding at segment boundaries. We further process the input using overlapping sliding windows with a stride of $s = \lfloor L_w(1-r) \rfloor$, where $L_w$ is the segment length and $r$ is the overlap ratio, accumulating per-class votes across all windows before taking the frame-wise argmax. 

The wider architecture design trades depth for parallel processing capacity: the 2E1D model generally runs faster than BTC. As shown in Section~\ref{sec:results}, the wider 2E1D architecture is more susceptible to degradation when adapting to noisy labels compared to the deeper architecture of BTC. Thus, the 2E1D model requires stronger KD regularization to maintain stability.

\begin{table*}
\centering
\scriptsize
\setlength{\tabcolsep}{2.4pt}
\renewcommand{\arraystretch}{1.05}

\begin{tabular}{llcccccccccccccc}
\toprule
& & \multicolumn{7}{c}{\texttt{mir\_eval} Metrics} & \multicolumn{3}{c}{Segmentation} & \multicolumn{4}{c}{Frame-wise vs. teacher} \\
\cmidrule(lr){3-9} \cmidrule(lr){10-12} \cmidrule(lr){13-16}
\textbf{Model} & \textbf{Training Data} & \textbf{Root} & \textbf{Thirds} & \textbf{Triads} & \textbf{7ths} & \textbf{Tetrads} & \textbf{Majmin} & \textbf{MIREX} & \textbf{Over} & \textbf{Under} & \textbf{Seg} & \textbf{Acc} & \textbf{Prec} & \textbf{Rec} & \textbf{F1} \\
% \midrule
% \multicolumn{16}{l}{\textit{Pre-trained teacher model (reference)}} \\
\midrule
BTC \cite{Park19} & Pre-trained weights & 81.89 & 78.49 & 76.85 & 66.29 & 63.72 & 79.00 & 78.65 & 82.16 & 90.33 & 80.85 & -- & -- & -- & -- \\
\midrule
\multicolumn{16}{l}{\textit{Our two student models trained only with pseudo-labels (generated by the BTC teacher from unlabeled data)}} \\
\midrule
\multirow{5}{*}{2E1D (2.2M)}
& FMA (short-form) & 77.28{\scriptsize$\pm$1.3} & 73.70{\scriptsize$\pm$1.2} & 72.15{\scriptsize$\pm$1.1} & 61.78{\scriptsize$\pm$1.4} & 58.94{\scriptsize$\pm$1.5} & 74.49{\scriptsize$\pm$1.2} & 73.68{\scriptsize$\pm$1.3} & 83.55 & 83.96 & 78.03 & 57.27 & 34.43 & 20.29 & 23.51 \\
& DALI (long-form) & 74.32{\scriptsize$\pm$1.1} & 65.53{\scriptsize$\pm$1.3} & 63.73{\scriptsize$\pm$1.4} & 52.68{\scriptsize$\pm$1.5} & 50.27{\scriptsize$\pm$1.4} & 65.91{\scriptsize$\pm$1.3} & 65.37{\scriptsize$\pm$1.4} & 77.09 & \cellcolor{e1dgreenhighlight}\textbf{85.92} & 74.00 & 78.25 & 48.06 & 37.41 & 39.52 \\
& MAESTRO+DALI & 79.50{\scriptsize$\pm$1.0} & 75.80{\scriptsize$\pm$1.0} & 74.11{\scriptsize$\pm$1.1} & 63.18{\scriptsize$\pm$1.3} & 60.28{\scriptsize$\pm$1.2} & 76.53{\scriptsize$\pm$1.0} & 75.50{\scriptsize$\pm$1.1} & 83.14 & 84.85 & 78.53 & 79.44 & 52.23 & 41.95 & 44.50 \\
& All (FMA+M+D) & \cellcolor{e1dgreenhighlight}\textbf{80.37{\scriptsize$\pm$1.2}} & \cellcolor{e1dgreenhighlight}\textbf{76.63{\scriptsize$\pm$1.1}} & \cellcolor{e1dgreenhighlight}\textbf{74.91{\scriptsize$\pm$1.2}} & \cellcolor{e1dgreenhighlight}\textbf{64.37{\scriptsize$\pm$1.4}} & \cellcolor{e1dgreenhighlight}\textbf{61.50{\scriptsize$\pm$1.3}} & \cellcolor{e1dgreenhighlight}\textbf{77.29{\scriptsize$\pm$1.1}} & \cellcolor{e1dgreenhighlight}\textbf{76.35{\scriptsize$\pm$1.2}} & \cellcolor{e1dgreenhighlight}\textbf{84.34} & 85.63 & \cellcolor{e1dgreenhighlight}\textbf{80.15} & 80.15 & 57.01 & 45.20 & 47.49 \\
& All + KD ($\alpha{=}0.3$) & 80.17{\scriptsize$\pm$1.0} & 76.36{\scriptsize$\pm$1.0} & 74.68{\scriptsize$\pm$1.0} & 63.93{\scriptsize$\pm$1.3} & 61.06{\scriptsize$\pm$1.4} & 77.09{\scriptsize$\pm$1.0} & 76.08{\scriptsize$\pm$1.0} & 84.23 & 85.27 & 79.69 & \cellcolor{e1dgreenhighlight}\textbf{81.23} & \cellcolor{e1dgreenhighlight}\textbf{58.14} & \cellcolor{e1dgreenhighlight}\textbf{46.33} & \cellcolor{e1dgreenhighlight}\textbf{48.62} \\
\midrule
\multirow{5}{*}{ BTC (3.03M) }
& FMA (short-form) & 78.85{\scriptsize$\pm$1.2} & 74.76{\scriptsize$\pm$1.0} & 73.12{\scriptsize$\pm$1.1} & 62.88{\scriptsize$\pm$1.3} & 60.09{\scriptsize$\pm$1.4} & 75.37{\scriptsize$\pm$1.0} & 75.07{\scriptsize$\pm$1.1} & 80.92 & 87.57 & 79.44 & 61.12 & 36.53 & 24.11 & 26.37 \\
& DALI (long-form) & 81.01{\scriptsize$\pm$1.0} & 77.24{\scriptsize$\pm$0.9} & 75.35{\scriptsize$\pm$1.0} & 65.46{\scriptsize$\pm$1.2} & 62.59{\scriptsize$\pm$1.1} & 77.66{\scriptsize$\pm$0.9} & 77.14{\scriptsize$\pm$1.0} & 79.50 & 90.60 & 79.51 & 85.88 & 49.23 & 50.37 & 47.32 \\
& MAESTRO+DALI & 80.65{\scriptsize$\pm$1.1} & 76.37{\scriptsize$\pm$1.0} & 74.61{\scriptsize$\pm$1.2} & 64.19{\scriptsize$\pm$1.0} & 61.35{\scriptsize$\pm$1.3} & 76.93{\scriptsize$\pm$0.9} & 76.69{\scriptsize$\pm$1.0} & 77.42 & \cellcolor{btcbluehighlight}\textbf{90.67} & 77.66 & 88.01 & 58.84 & \cellcolor{btcbluehighlight}\textbf{64.29} & 59.16 \\
& All (FMA+M+D) & \cellcolor{btcbluehighlight}\textbf{81.54{\scriptsize$\pm$0.9}} & \cellcolor{btcbluehighlight}\textbf{77.86{\scriptsize$\pm$0.7}} & \cellcolor{btcbluehighlight}\textbf{76.08{\scriptsize$\pm$0.8}} & \cellcolor{btcbluehighlight}\textbf{66.29{\scriptsize$\pm$1.1}} & \cellcolor{btcbluehighlight}\textbf{63.54{\scriptsize$\pm$1.0}} & \cellcolor{btcbluehighlight}\textbf{78.29{\scriptsize$\pm$0.7}} & \cellcolor{btcbluehighlight}\textbf{77.84{\scriptsize$\pm$0.9}} & \cellcolor{btcbluehighlight}\textbf{80.97} & 90.15 & \cellcolor{btcbluehighlight}\textbf{80.76} & 89.14 & 62.54 & 63.89 & 59.34 \\
& All + KD ($\alpha{=}0.3$) & 81.23{\scriptsize$\pm$1.0} & 77.75{\scriptsize$\pm$0.8} & 75.94{\scriptsize$\pm$0.9} & 66.00{\scriptsize$\pm$1.0} & 63.22{\scriptsize$\pm$1.1} & 78.15{\scriptsize$\pm$0.8} & 77.82{\scriptsize$\pm$1.0} & 80.81 & 90.25 & 80.60 & \cellcolor{btcbluehighlight}\textbf{89.47} & \cellcolor{btcbluehighlight}\textbf{63.21} & 64.05 & \cellcolor{btcbluehighlight}\textbf{60.12} \\
\bottomrule
\end{tabular}
\caption{Pseudo-labeling results across dataset configurations. The \texttt{mir\_eval} metrics are computed against the ground-truth test set, while frame-wise accuracy, precision, recall, and F1 measure agreement with the teacher's predictions. FMA contains short-form clips ($\sim$30s), while DALI and MAESTRO provide long-form full tracks. The pre-trained BTC teacher \cite{Park19} serves as the reference/upper bound on \texttt{mir\_eval} metrics. Best results for BTC and 2E1D student models in the table are highlighted in blue and green, respectively.}
\label{tab:comprehensive}
\end{table*}

\section{Experiments}
\label{sec:experiment}

\subsection{Datasets and Preprocessing}
\label{sec:data}
\textbf{Unlabeled Datasets.} We use three large-scale unlabeled datasets for pseudo-label generation, totaling over 1,000 hours of audio. The Free Music Archive \cite{Defferrard17} ($\mathcal{D}_{FMA}$) provides over 100,000 short-form tracks ($\sim$30s) with extensive genre diversity. The MAESTRO dataset \cite{Hawthorne19} ($\mathcal{D}_{MAESTRO}$) contributes approximately 200 hours of high-quality piano recordings with precise MIDI alignment. The DALI dataset \cite{Meseguer-Brocal2018} ($\mathcal{D}_{DALI}$) supplies over 5,000 full-length music tracks.\\
\textbf{Labeled Datasets.} We aggregate annotations from the Isophonics \cite{Mauch09}, the McGill Billboard corpus \cite{Burgoyne11}, the RWC Pop dataset \cite{Goto02}, and the USPop annotations distributed with \cite{McFee17}, collecting a fixed subset of 600 songs. The dataset is then split in a ratio of 7:1:2 into train/validation/test sets (420/60/120 songs). The ``50\%'' and ``full'' conditions in our continual learning experiments refer to using 210 and 420 training songs from this subset, respectively.\\
\textbf{Clean vs. Noisy Annotations.} To evaluate KD's robustness to annotation noise, we prepare two versions of the labeled data: (1) \textit{clean labels} with careful manual alignment, and (2) \textit{noisy labels} sourced online without alignment correction, which primarily affect non-chord (``N'') label boundaries. This setup enables an ablation study of KD's regularization effect under different noise conditions.\\
\textbf{Preprocessing.} All audio undergoes identical preprocessing. The CQT features \cite{Schoerkhuber10} are extracted with $F=144$ frequency bins, 24 bins per octave, and a hop length of $h=2048$ samples, yielding a temporal resolution of $\Delta t \approx \SI{93}{\milli\second}$ at sampling rate $f_{sr} = \SI{22.05}{\kilo\hertz}$. Features undergo z-score normalization using teacher model statistics ($\mu_t$, $\sigma_t$) to maintain identical input distributions. 

\subsection{Training Configuration}
\label{sec:config}
In the first stage, pseudo-labeling training employs the AdamW optimizer with a batch size of 256 and a sequence length of 108 frames ($\sim$10 seconds). The learning rate is warmed up from $10^{-4}$ to a peak of $3\cdot10^{-4}$ over 10 epochs, followed by cosine-annealed decay. Early stopping monitors validation accuracy with a patience of 10 epochs. We reserve 10\% of the pseudo-labeled data for validation and 10\% as a held-out test set. In the second stage, continual learning, we use a reduced learning rate of $10^{-5}$ with decay upon observing a validation plateau. KD weight $\alpha{=}0.3$ balances adaptation and regularization. The temperature $\tau{=}3.0$ is empirically selected as the optimal value in all settings. Training continues until early stopping triggers. No data augmentation is applied in our two-stage pipeline (Section~\ref{sec:augmentation}).

For the supervised learning (SL) baseline, both BTC and 2E1D are trained from scratch on the full labeled training set (420 songs). The SL baseline uses the same configuration as in pseudo-labeling training, except for the learning schedule. The learning rate decays
by a factor of 0.9 when validation accuracy does not improve. Pitch-shifting augmentation via the Rubber Band library transposes both audio and labels from $-5$ to $+6$ semitones.

\subsection{Evaluation Metrics}
\label{sec:metrics}
We employ standard metrics from the \texttt{mir\_eval} library \cite{raffel14}. The Root metric compares the root note. The Thirds metric adds major and minor third intervals. The Triads metric evaluates all triadic qualities, while Majmin focuses on major and minor qualities. The Sevenths metric measures a predefined set of seventh chords. The Tetrads metric extends evaluation to four tones. MIREX considers an estimation accurate when at least three pitch classes are correct. Frame-wise accuracy (Acc), precision (Prec), recall (Rec), and F1 are computed using standard True Positive, False Positive, and False Negative counts. Additionally, we report Chord Symbol Recall (CSR) and Weighted Chord Symbol Recall (WCSR) following \cite{Bortolozzo21}:
\begin{equation}
\CSR_{c,i} = \frac{|S_{c,i}^{\mathrm{pred}} \cap S_{c,i}^{\mathrm{ref}}|}{|S_{c,i}^{\mathrm{ref}}|}, \quad \WCSR_c = \frac{\sum_i T_i \cdot \CSR_{c,i}}{\sum_i T_i},
\end{equation}
where, for a chord class $c$ and track index $i$, $S_{c,i}^{\mathrm{pred}}$ denotes predicted chord labels, $S_{c,i}^{\mathrm{ref}}$ denotes ground-truth labels for the respective intervals, and $T_i$ is the duration of the $i$-th track. The Average Chord Quality Accuracy (ACQA), with the chord set $\mathcal{V}$, is calculated as:
\begin{equation}
\ACQA = \frac{\sum_{c\in \mathcal{V}} \WCSR_{c}}{|\mathcal{V}|}.
\end{equation}
WCSR weights accuracy by class distribution, while ACQA gives equal weight to all chord qualities. This makes ACQA more sensitive to rare chord performance. Finally, we report segmentation metrics using the standard library implementation~\cite{raffel14}. The over-segmentation score (Over) measures how well estimated boundaries avoid fragmenting the ground-truth reference segments, while the under-segmentation score (Under) measures how well predicted segments avoid merging them; higher values indicate better boundary agreement. The overall segmentation score (Seg) is the minimum of the two scores, macro-averaged across tracks.

\begin{table*}
\centering
\scriptsize
\setlength{\tabcolsep}{4.5pt}
\begin{tabular}{lcccccccccc}
\toprule

\textbf{Method} & \textbf{Root} & \textbf{Thirds} & \textbf{Triads} & \textbf{7ths} & \textbf{Tetrads} & \textbf{Majmin} & \textbf{MIREX} & \textbf{Seg} & \textbf{WCSR} & \textbf{ACQA} \\
\midrule
\multicolumn{11}{l}{\textit{Prior Work (Noisy Student Framework)}} \\
\midrule
Bortolozzo et al. \cite{Bortolozzo21} \textsuperscript{\textdagger}  & -- & -- & -- & -- & -- & -- & -- & -- & 46.8 & 24.5 \\
Li et al. \cite{Li23_contrastive} \textsuperscript{\textdagger}  & 78.05 & 73.56 & 70.71 & 61.11 & 56.84 & 74.64 & 74.26 & -- & -- & -- \\
\midrule
\multicolumn{11}{l}{\textit{Traditional supervised learning (SL) baseline (trained on full labels with data augmentation)}} \\
\midrule
2E1D (SL) (2.2M) & 79.39{\scriptsize$\pm$1.4} & 76.14{\scriptsize$\pm$1.2} & 74.53{\scriptsize$\pm$0.6} & 64.53{\scriptsize$\pm$1.9} & 61.67{\scriptsize$\pm$1.0} & 76.89{\scriptsize$\pm$1.4} & 76.24{\scriptsize$\pm$0.7} & 78.50{\scriptsize$\pm$1.2} & 68.7{\scriptsize$\pm$1.0} & 24.4{\scriptsize$\pm$1.5} \\
BTC (SL) (3.03M) & 81.52{\scriptsize$\pm$1.2} & 78.00{\scriptsize$\pm$1.0} & 76.12{\scriptsize$\pm$0.5} & 65.93{\scriptsize$\pm$1.7} & 63.44{\scriptsize$\pm$0.9} & 78.11{\scriptsize$\pm$1.2} & 77.79{\scriptsize$\pm$0.9} & 79.38{\scriptsize$\pm$1.1} & 68.6{\scriptsize$\pm$1.1} & 29.0{\scriptsize$\pm$1.3} \\
\midrule
\multicolumn{11}{l}{\textit{Results of our student models after Stage 2 with Data-Incremental Continual Learning (with clean labels, $\alpha{=}0.3$)}} \\
\midrule
2E1D (CL) (50\%) & 81.02{\scriptsize$\pm$0.08} & 77.54{\scriptsize$\pm$0.18} & 75.76{\scriptsize$\pm$0.20} & 66.39{\scriptsize$\pm$0.18} & 63.68{\scriptsize$\pm$0.23} & 77.93{\scriptsize$\pm$0.17} & 77.57{\scriptsize$\pm$0.09} & 80.23{\scriptsize$\pm$0.21} & 70.4{\scriptsize$\pm$0.3} & 34.2{\scriptsize$\pm$0.6} \\
2E1D (CL) (full) & \cellcolor{e1dgreenhighlight}\textbf{81.55{\scriptsize$\pm$0.05}} & \cellcolor{e1dgreenhighlight}\textbf{78.52{\scriptsize$\pm$0.12}} & \cellcolor{e1dgreenhighlight}\textbf{76.85{\scriptsize$\pm$0.13}} & \cellcolor{e1dgreenhighlight}\textbf{68.19{\scriptsize$\pm$0.12}} & \cellcolor{e1dgreenhighlight}\textbf{65.49{\scriptsize$\pm$0.15}} & \cellcolor{e1dgreenhighlight}\textbf{78.98{\scriptsize$\pm$0.11}} & \cellcolor{e1dgreenhighlight}\textbf{78.52{\scriptsize$\pm$0.06}} & \cellcolor{e1dgreenhighlight}\textbf{80.73{\scriptsize$\pm$0.14}} & \cellcolor{e1dgreenhighlight}\textbf{71.9{\scriptsize$\pm$0.2}} & \cellcolor{e1dgreenhighlight}\textbf{36.1{\scriptsize$\pm$0.4}} \\
BTC (CL) (50\%) & 82.14{\scriptsize$\pm$0.06} & 78.58{\scriptsize$\pm$0.17} & 76.67{\scriptsize$\pm$0.18} & 66.77{\scriptsize$\pm$0.15} & 64.28{\scriptsize$\pm$0.20} & 78.62{\scriptsize$\pm$0.15} & 78.87{\scriptsize$\pm$0.08} & 81.17{\scriptsize$\pm$0.18} & 68.7{\scriptsize$\pm$0.2} & 37.1{\scriptsize$\pm$0.6} \\
BTC (CL) (full) & \cellcolor{btcbluehighlight}\textbf{83.03{\scriptsize$\pm$0.04}} & \cellcolor{btcbluehighlight}\textbf{80.17{\scriptsize$\pm$0.11}} & \cellcolor{btcbluehighlight}\textbf{78.33{\scriptsize$\pm$0.12}} & \cellcolor{btcbluehighlight}\textbf{69.38{\scriptsize$\pm$0.10}} & \cellcolor{btcbluehighlight}\textbf{67.00{\scriptsize$\pm$0.13}} & \cellcolor{btcbluehighlight}\textbf{80.24{\scriptsize$\pm$0.10}} & \cellcolor{btcbluehighlight}\textbf{80.16{\scriptsize$\pm$0.05}} & \cellcolor{btcbluehighlight}\textbf{81.71{\scriptsize$\pm$0.12}} & \cellcolor{btcbluehighlight}\textbf{71.6{\scriptsize$\pm$0.1}} & \cellcolor{btcbluehighlight}\textbf{39.5{\scriptsize$\pm$0.4}} \\
\bottomrule
\end{tabular}
\caption{Data-incremental continual learning results with overall evaluation metrics. After pseudo-labeling, we fine-tune on ground-truth labels with KD ($\alpha{=}0.3$). Results for the student models are reported as the average and standard deviation (Mean $\pm$ Std). Best results for BTC and 2E1D are highlighted in blue and green, respectively. \textsuperscript{\textdagger} Evaluated on our test split using our re-implementation (reporting only the metrics compatible with the model's prediction vocabulary).}

\label{tab:continual}
\end{table*}

\section{Results}
\label{sec:results}
In this section, we present results for our two-stage training method across the metrics described in Section~\ref{sec:metrics}, including an ablation study on the KD regularization against noisy labels. All reported metrics are evaluated on the held-out test set (120 songs) from the clean labeled dataset described in Section~\ref{sec:data}.

\subsection{Stage 1: Training with pseudo-labels}
Increasing unlabeled data diversity consistently improves performance across the \texttt{mir\_eval} metric hierarchy and segmentation quality (Table~\ref{tab:comprehensive}). Training on all three datasets (FMA, MAESTRO, DALI) yields the strongest results for both architectures, indicating that pseudo-label pre-training benefits from coverage of varied genres and recording conditions. Long-form datasets consisting of full-length tracks (DALI, MAESTRO) produce more stable pseudo-labels than short-form clips (FMA): DALI achieves higher frame-wise agreement with the teacher than FMA. Longer musical contexts provide consistent harmonic progressions, reducing boundary jitter and spurious chord predictions. However, FMA achieves better \texttt{mir\_eval} results than DALI for 2E1D due to its larger size. The best BTC student reaches about 99\% of teacher performance on ground-truth evaluation across all \texttt{mir\_eval} metrics, while the purely transformer-based 2E1D achieves 96--98\% of the teacher model results, demonstrating cross-architecture knowledge transfer. When all training targets are pseudo-labels, adding KD better aligns the student to the teacher’s soft-label distribution, resulting in improvements across frame-wise metrics despite a decrease in \texttt{mir\_eval} metrics. In addition, KD accelerates optimization, as all pseudo-label training runs using KD converge in 30–40 epochs versus 50–70 for other settings; this comes with a modest training-time memory overhead to store and backpropagate full soft targets.
%\\  Demo: \url{https://www.youtube.com/watch?v=IFybAITJuCk}}
%\footnote{A chord recognition application using the trained model is available at: \url{https://github.com/anonymous/repo-app}. \\ 
%Demo: \url{https://www.youtube.com/watch?v=youtube-id}}

\subsection{Stage 2: Continually training with GT + KD}

In Stage 2, the student weights from Stage 1 pseudo-label pre-training are continually trained on ground-truth labeled data with selective KD ($\alpha{=}0.3$). BTC with full labels consistently surpasses both the teacher (Table~\ref{tab:comprehensive}) and the supervised learning (SL) baseline across all seven \texttt{mir\_eval} metrics (Table~\ref{tab:continual}). The 2E1D student similarly improves over its SL baseline across the board. Among the recognition hierarchy, 7ths and Tetrads benefit the most, indicating that combining broad pseudo-label pre-training with targeted ground-truth fine-tuning is particularly effective for complex chords. For comparison with prior semi-supervised approaches~\cite{Bortolozzo21, Li23_contrastive}, we re-implemented their training procedures as described in the respective original publications and evaluated the resulting models on our held-out test split under identical preprocessing settings. Prior work relies on weaker supervised models to generate pseudo-labels, constraining label quality; in contrast, leveraging a highly capable teacher in Stage 1 provides a stronger initialization for Stage 2 adaptation. Furthermore, we observe a substantial drop in seed-to-seed variance in Stage 2 ($\pm 0.04$ to $0.23$) compared to Stage 1 ($\pm 0.7$ to $1.5$) on the \texttt{mir\_eval} metrics. This stability arises because Stage 2 optimizes over unambiguous ground-truth labels rather than noisy pseudo-labels, while selective KD further acts as a regularizer to prevent the model from overfitting to noise in the smaller labeled dataset.

Notably, the BTC student model is capable of surpassing the teacher model that generated its pre-training pseudo-labels. We attribute this to two complementary mechanisms. First, in Stage 1, the student is trained on a substantially larger unlabeled dataset than the labeled data available to the teacher. The larger amount of data allows the student to approximate the teacher’s underlying knowledge more accurately. Second, in Stage 2, ground-truth annotations allow the student to correct systematic errors inherited from the teacher, particularly on rare chord qualities where the teacher's predictions are least reliable. Additionally, selective KD simultaneously anchors the student to the teacher's well-calibrated representations on common chords, preventing overfitting to the small labeled set. The combination of broader pre-training coverage and targeted error correction thus enables the student to exceed the teacher's overall performance. 

\begin{figure}[htb]
    \centering
    \includegraphics[width=0.48\textwidth]{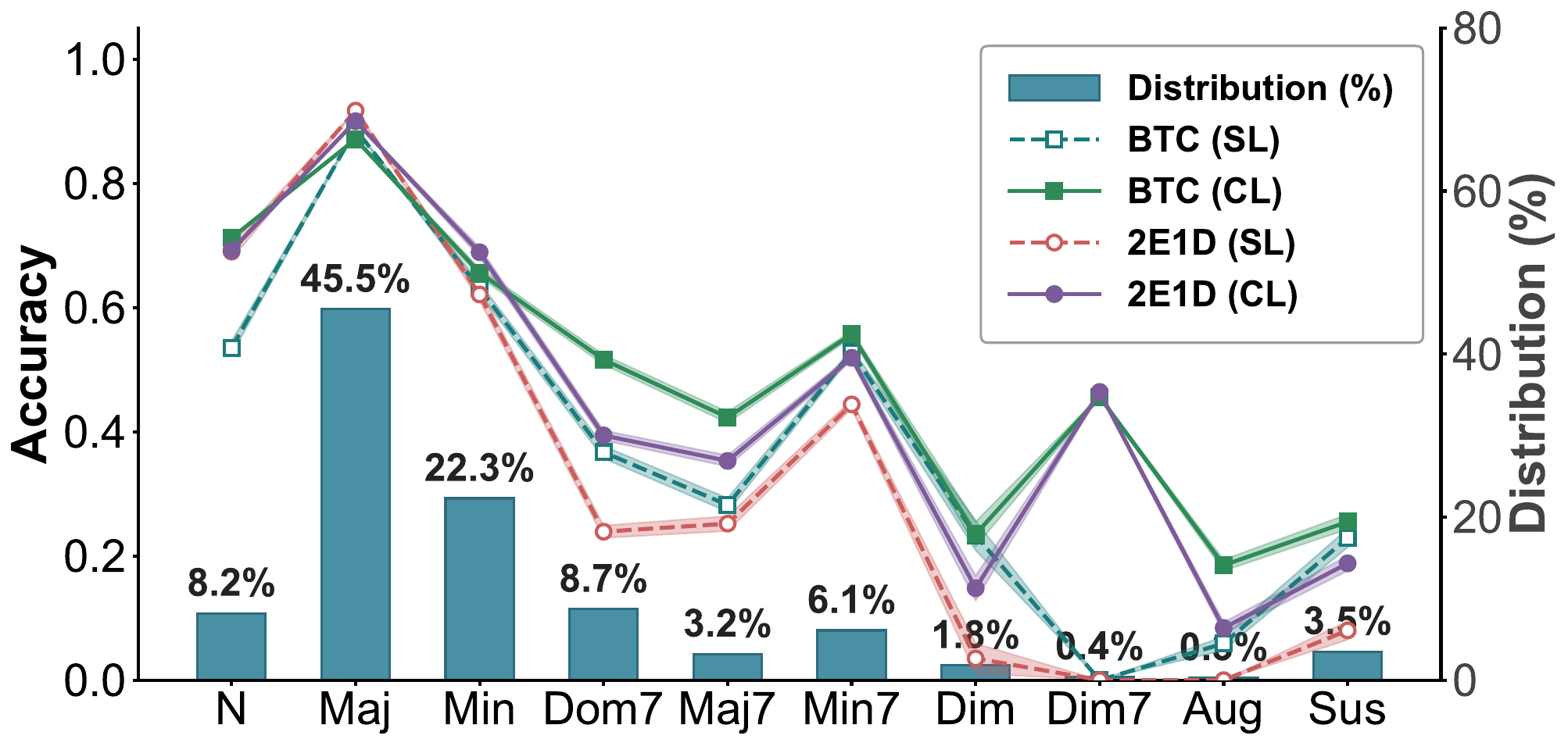}
    \caption{Chord quality distribution (bars) and recognition accuracy (lines) of models trained with supervised learning (BTC (SL), 2E1D (SL)) versus our approach (BTC (CL), 2E1D (CL)).}
    \label{fig:chord_dist}
\end{figure}

Figure~\ref{fig:chord_dist} provides a per-quality view of this improvement and serves as a central illustration of our approach. The training distribution is heavily imbalanced: Major and Minor chords dominate the training frames, while rare qualities (Dim, Dim7, Aug, Sus) constitute less than 3\%. The SL baselines achieve 0\% accuracy on Dim7 and 0\% (2E1D) or 5.9\% (BTC) on Aug due to insufficient training examples; our pipeline yields measurable improvements on these classes. The absolute test-sample counts reported in Table~\ref{tab:subset_quality} (e.g., 212 frames for Dim7, 517 for Aug) confirm that these gains reflect consistent frame-level corrections rather than statistical noise. The ACQA metric, which weights all qualities equally, captures this disproportionate rare-chord improvement more clearly than distribution-weighted WCSR. The pre-training on pseudo-labels also explains the sample-efficiency effect: the student trained on only 50\% of labeled data already outperforms the supervised baseline trained on the full dataset (Table~\ref{tab:continual}).

\begin{table}[htb]
\centering
\scriptsize
\setlength{\tabcolsep}{1.7pt}
\begin{tabular}{lcccccccccc}
  \toprule
  \textbf{Method} & \textbf{N} & \textbf{Maj} & \textbf{Min} & \textbf{Dom7} & \textbf{Maj7} & \textbf{Min7} & \textbf{Dim} & \textbf{Dim7} & \textbf{Aug} & \textbf{Sus} \\
  \multicolumn{1}{r}{\textbf{Count}} & 14.7k & 137k & 36.4k & 32.6k & 11.4k & 25.3k & 540 & 212 & 517 & 5.6k \\
  \midrule
  Bortolozzo et al.~\cite{Bortolozzo21}\textsuperscript{\textdagger} & -- & 55.5 & 54.4 & 51.0 & 12.0 & 47.5 & 30.4 & 43.3 & -- & -- \\
  \midrule
  2E1D (SL) & 69.0 & \cellcolor{e1dgreenhighlight}\textbf{91.7} & 62.1 & 23.9 & 25.2 & 44.4 & 3.5 & 0.0 & 0.0 & 8.0 \\
  BTC (SL) & 53.5 & \cellcolor{btcbluehighlight}\textbf{88.7} & 63.1 & 36.7 & 28.2 & 52.8 & 23.3 & 0.0 & 5.9 & 22.9 \\
  \midrule
  2E1D (CL) (50\%) & 69.2 & 90.9 & 63.0 & 33.3 & 29.5 & 49.0 & 12.7 & \cellcolor{e1dgreenhighlight}\textbf{46.7} & 5.2 & 16.6 \\
  2E1D (CL) (full) & \cellcolor{e1dgreenhighlight}\textbf{69.2} & 90.0 & \cellcolor{e1dgreenhighlight}\textbf{68.9} & \cellcolor{e1dgreenhighlight}\textbf{39.4} & \cellcolor{e1dgreenhighlight}\textbf{35.3} & \cellcolor{e1dgreenhighlight}\textbf{51.9} & \cellcolor{e1dgreenhighlight}\textbf{14.8} & 46.4 & \cellcolor{e1dgreenhighlight}\textbf{8.4} & \cellcolor{e1dgreenhighlight}\textbf{18.8} \\
  BTC (CL) (50\%) & 71.2 & 84.5 & 61.2 & 50.2 & 42.1 & 51.6 & 18.3 & \cellcolor{btcbluehighlight}\textbf{47.4} & 14.8 & 22.3 \\
  BTC (CL) (full) & \cellcolor{btcbluehighlight}\textbf{71.2} & 87.1 & \cellcolor{btcbluehighlight}\textbf{65.6} & \cellcolor{btcbluehighlight}\textbf{51.6} & \cellcolor{btcbluehighlight}\textbf{42.3} & \cellcolor{btcbluehighlight}\textbf{55.7} & \cellcolor{btcbluehighlight}\textbf{23.5} & 45.6 & \cellcolor{btcbluehighlight}\textbf{18.5} & \cellcolor{btcbluehighlight}\textbf{25.5} \\
  \bottomrule
\end{tabular}
\caption{Subset chord quality accuracy results after Stage 2 with Data-Incremental Continual Learning ($\alpha{=}0.3$). Best results for BTC and 2E1D are highlighted in blue and green, respectively. \textsuperscript{\textdagger} Evaluated on our test split using our re-implementation.}

\label{tab:subset_quality}
\end{table}

\begin{figure}[htb]
    \centering
    \includegraphics[width=0.5\textwidth]{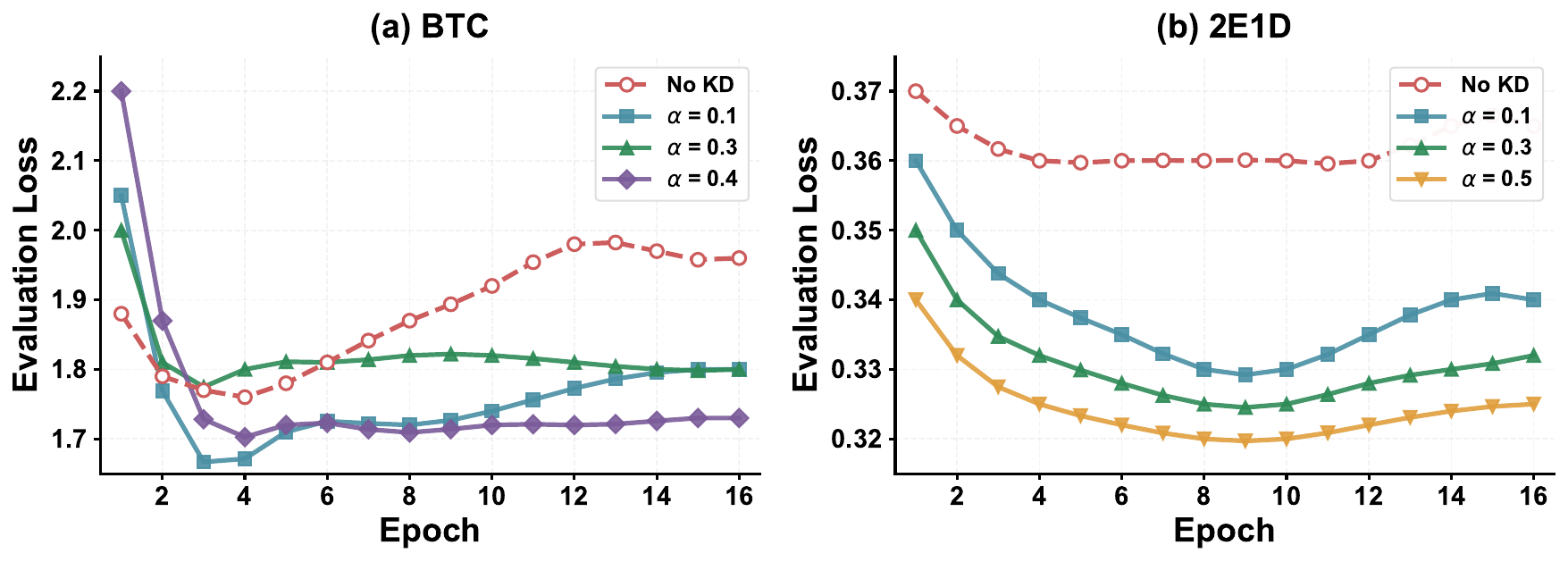}
    \caption{Evaluation loss of student models during continual training with different KD weights $\alpha$.}
    \label{fig:kd_effects}
\end{figure}

\textbf{KD as regularization under label noise.}
The above results use clean, manually aligned ground-truth labels. To isolate KD's role as a regularizer, we repeat Stage 2 using the same Stage 1 initialized weights but with noisy labels sourced online without alignment correction, which primarily affect non-chord (``N'') label boundaries. Table~\ref{tab:kd_effect} reports results across varying $\alpha$ values. Without KD ($\alpha{=}0$), both architectures degrade substantially: BTC drops consistently across all metrics, while the wider 2E1D collapses more severely. Increasing $\alpha$ progressively recovers performance; BTC peaks at $\alpha{=}0.3$, while 2E1D requires stronger regularization ($\alpha{=}0.5$) to stabilize. Figure~\ref{fig:kd_effects} shows the corresponding training dynamics: without KD, validation loss rises as the model overfits to noisy labels, whereas $\alpha > 0$ anchors the student to the teacher's distribution, preserving the pseudo-label knowledge acquired in Stage 1. Crucially, when labels are clean (Table~\ref{tab:continual}), KD does not impede adaptation. This confirms that KD selectively mitigates noise while preserving adaptation capacity.

\begin{table}[hbt!]
\centering
\scriptsize
\setlength{\tabcolsep}{2.5pt}
\begin{tabular}{lccccccccc}
\toprule
\textbf{Model} & \textbf{$\alpha$} & \textbf{Root} & \textbf{Thirds} & \textbf{Triads} & \textbf{7ths} & \textbf{Tetrads} & \textbf{Majmin} & \textbf{MIREX} & \textbf{Seg} \\
\midrule
\multirow{4}{*}{2E1D}
& 0 & 52.47 & 50.70 & 49.88 & 42.52 & 41.00 & 51.12 & 50.99 & 66.72 \\
& 0.1 & 55.85 & 54.16 & 53.41 & 47.68 & 45.78 & 54.88 & 54.07 & 67.70 \\
& 0.3 & 66.05 & 63.89 & 62.79 & 55.78 & 53.39 & 64.70 & 63.87 & 73.32 \\
& 0.5 & \cellcolor{e1dgreenhighlight}\textbf{74.21} & \cellcolor{e1dgreenhighlight}\textbf{71.25} & \cellcolor{e1dgreenhighlight}\textbf{69.78} & \cellcolor{e1dgreenhighlight}\textbf{61.20} & \cellcolor{e1dgreenhighlight}\textbf{58.35} & \cellcolor{e1dgreenhighlight}\textbf{72.09} & \cellcolor{e1dgreenhighlight}\textbf{71.54} & \cellcolor{e1dgreenhighlight}\textbf{76.84} \\
\midrule
\multirow{4}{*}{BTC}
& 0 & 71.06 & 68.05 & 66.73 & 58.11 & 55.66 & 68.69 & 68.20 & 77.05 \\
& 0.1 & 73.88 & 70.86 & 69.43 & 60.91 & 58.25 & 71.52 & 70.96 & 78.59 \\
& 0.3 & \cellcolor{btcbluehighlight}\textbf{77.34} & \cellcolor{btcbluehighlight}\textbf{73.95} & \cellcolor{btcbluehighlight}\textbf{72.44} & \cellcolor{btcbluehighlight}\textbf{62.97} & \cellcolor{btcbluehighlight}\textbf{60.25} & \cellcolor{btcbluehighlight}\textbf{74.60} & \cellcolor{btcbluehighlight}\textbf{74.29} & \cellcolor{btcbluehighlight}\textbf{79.61} \\
& 0.4 & 76.95 & 72.88 & 71.30 & 61.30 & 58.57 & 73.59 & 73.18 & 78.98 \\
\bottomrule
\end{tabular}
\caption{KD regularization effect during continual training with misaligned labels. The KD weight $\alpha$ controls the balance between teacher soft targets and ground-truth hard labels. Best results for BTC and 2E1D are highlighted in blue and green, respectively.}
\label{tab:kd_effect}
\end{table}

\section{Conclusion}
\label{sec:conclusion}

Since model weights are often more readily available than proprietary training data, we present a practical training strategy for the ACR problem that leverages open-weight pre-trained models when high-quality labels are scarce. We show that students trained solely on pseudo-labels can approach teacher-level performance across seven \texttt{mir\_eval} metrics. We further demonstrate that continual learning can improve performance without catastrophic forgetting when the teacher provides sufficiently general representations. Under our pipeline, the best student model ultimately surpasses the teacher, with major gains on rare chord qualities (e.g., Dim, Dim7, Aug). Knowledge distillation improves robustness to noisy labels while preserving adaptability when ground-truth annotations are clean. We also find that the wider 2E1D architecture requires stronger KD regularization than the deeper BTC, underscoring how architectural choices influence continual-learning stability. A key limitation is reliance on teacher quality: biased or weakly generalizable teacher representations can transfer these shortcomings to the student. Future work will explore stronger teacher models and model architectures, ensembles of multiple teachers, scaling to additional unlabeled corpora, and extending the framework to related MIR tasks such as beat tracking and key estimation.

%\newpage
\bibliographystyle{IEEEtranDAFx}
\bibliography{DAFx26_tmpl}

\end{document}